\documentclass[11pt, conference]{IEEEtran}
\makeatletter
\def\ps@headings{%
\def\@oddhead{\mbox{}\scriptsize\rightmark \hfil \thepage}%
\def\@evenhead{\scriptsize\thepage \hfil \leftmark\mbox{}}%
\def\@oddfoot{}%
\def\@evenfoot{}}
\makeatother 
\ifCLASSINFOpdf
\else
\usepackage[dvips]{graphicx}
\fi

\usepackage[font=footnotesize]{subfig}

\usepackage{times}
\usepackage{epsfig}
\usepackage{graphicx}
\begin{document}
%
\title{Rhythm and Randomness in Human Contact}

\author{
\IEEEauthorblockN{Mervyn P. Freeman, Nicholas W. Watkins}
\IEEEauthorblockA{British Antarctic Survey\\
Cambridge, United Kingdom\\
Email: \{mpf,~nww\}@bas.ac.uk}
\and
\IEEEauthorblockN{Eiko Yoneki, Jon Crowcroft}
\IEEEauthorblockA{University of Cambridge\\
Cambridge, United Kingdom\\
Email: \{eiko.yoneki,~jon.crowcroft\}@cl.cam.ac.uk} }

\maketitle

\begin{abstract}
There is substantial interest in the effect of human mobility patterns on opportunistic
communications. Inspired by recent work revisiting some of the early evidence for a L\'{e}vy flight
foraging strategy in animals, we analyse datasets on human contact from  real world traces. By
analysing the distribution of inter-contact times on different time scales and using different
graphical forms, we find not only the highly skewed distributions of waiting times highlighted in
previous studies but also clear circadian rhythm. The relative visibility of these two components
depends strongly on which graphical form is adopted and the range of time scales. We use a simple
model to reconstruct the observed behaviour and discuss the implications of this for forwarding
efficiency.
\end{abstract}
\IEEEpeerreviewmaketitle
\begin{table*}[htbp]
\begin{center}
\begin{tabular}{|p{64pt}|p{49pt}|p{54pt}|p{48pt}|p{46pt}|p{49pt}|p{46pt}|p{57pt}|}
\hline
\textbf{~~}&
\textbf{~~}&
\textbf{~~}&
\textbf{~~}&
\textbf{~~}&
\textbf{~~}&
\textbf{~~}&
\textbf{~~} \\
\textbf{User population}& \textbf{Intel}& \textbf{Cambridge 1}& \textbf{INFOCOM 2005}&
\textbf{Toronto}& \textbf{UCSD}& \textbf{Dartmouth}&
\textbf{Europe} \\
\hline
Source&
\multicolumn{6}{|p{335pt}|}{\centering Chaintreau et al. (2006) \cite{psn-HUMAN}} &
Gonzalez et al. (2009) \cite{marta}\\
\hline
\textbf{~~}&
\textbf{~~}&
\textbf{~~}&
\textbf{~~}&
\textbf{~~}&
\textbf{~~}&
\textbf{~~}&
\textbf{~~} \\
Device&
iMote&
iMote&
iMote&
PDA&
PDA&
Laptop/PDA&
Mobile phone \\
\hline
\textbf{~~}&
\textbf{~~}&
\textbf{~~}&
\textbf{~~}&
\textbf{~~}&
\textbf{~~}&
\textbf{~~}&
\textbf{~~} \\
Network type&
Bluetooth&
Bluetooth&
Bluetooth&
Bluetooth&
WiFi&
WiFi&
Mobile phone \\
\hline
\textbf{~~}&
\textbf{~~}&
\textbf{~~}&
\textbf{~~}&
\textbf{~~}&
\textbf{~~}&
\textbf{~~}&
\textbf{~~} \\
Granularity&
120 seconds&
120 seconds&
120 seconds&
120 seconds&
120 seconds&
300 seconds&
N/A \\
\hline
\textbf{~~}&
\textbf{~~}&
\textbf{~~}&
\textbf{~~}&
\textbf{~~}&
\textbf{~~}&
\textbf{~~}&
\textbf{~~} \\
Duration&
3 days&
5 days&
3 days&
16 days&
77 days&
114 days&
6 months \\
\hline
\textbf{~~}&
\textbf{~~}&
\textbf{~~}&
\textbf{~~}&
\textbf{~~}&
\textbf{~~}&
\textbf{~~}&
\textbf{~~} \\
Devices participating&
8&
12&
41&
23&
273&
6648&
100,000 \\
\hline
\textbf{~~}&
\textbf{~~}&
\textbf{~~}&
\textbf{~~}&
\textbf{~~}&
\textbf{~~}&
\textbf{~~}&
\textbf{~~} \\
Number of internal contacts&
1,091&
4,229&
22,459&
2,802&
195,364&
4,058,284&
16,364,308 \\
\hline
\textbf{~~}&
\textbf{~~}&
\textbf{~~}&
\textbf{~~}&
\textbf{~~}&
\textbf{~~}&
\textbf{~~}&
\textbf{~~} \\
Approximate extent of power law region&
4min - 14min&
10min - 30min&
10min - 10h&
2min - 6min&
20min - 1day&
10min - 1h&
100s - 8h \\
\hline
\textbf{~~}&
\textbf{~~}&
\textbf{~~}&
\textbf{~~}&
\textbf{~~}&
\textbf{~~}&
\textbf{~~}&
\textbf{~~} \\
Quoted power law exponent&
-0.9&
-0.9&
-0.4&
-0.9&
-0.3&
-0.3&
-0.9 +/- 0.1 \\
\hline
\textbf{~~}&
\textbf{~~}&
\textbf{~~}&
\textbf{~~}&
\textbf{~~}&
\textbf{~~}&
\textbf{~~}&
\textbf{~~} \\
Type of distribution plotted&
Tail df \par(ccdf)&
Tail df \par(ccdf)&
Tail df \par(ccdf)&
Tail df \par(ccdf)&
Tail df \par(ccdf)&
Tail df \par(ccdf)&
Log-binned pdf \\
\hline
\textbf{~~}&
\textbf{~~}&
\textbf{~~}&
\textbf{~~}&
\textbf{~~}&
\textbf{~~}&
\textbf{~~}&
\textbf{~~} \\
\textbf{Inferred stability exponent $\alpha$}&
\textbf{0.9}&
\textbf{0.9}&
\textbf{0.4}&
\textbf{0.9}&
\textbf{0.3}&
\textbf{0.3}&
\textbf{-0.1 +/- 0.1} \\
\hline
\end{tabular}
{\bf \caption{{\bf Summary of studies in which the stability exponent
$\alpha$ has been inferred from an inter-contact time distribution}}}
\label{table:one}
\end{center}
\vspace{-7mm}
\end{table*}
\section{Introduction}
\noindent
Digital traffic flows not only over the wired backbone of the Internet or network of mobile phone
masts, but also in small leaps through physical space as people pass one another on the street
\cite{Kleinbergnature}. Thus opportunities for a new communication paradigm via wireless-enabled
devices are emerging, which communicate directly with other devices within their range and without
a costly and inflexible planned infrastructure (e.g., \cite{haggle}). To improve communication
efficiency and prevent the spread of wireless viruses in this new generation of communication
requires new insights and quantitative models of human interaction. Of fundamental importance in
this case is the time sequence of human contacts, as well as other properties of complex networks,
such as small-worldness, etc. (e.g., the special issue of Science on Complex Systems and Networks,
July 24, 2009).

Recently, the emergence of human interaction traces from online and pervasive environments is allowing us to
understand details of human activities. For example, the MIT Reality Mining project
{\cite{realityMining}} collected proximity, location and activity information, with nearby nodes
being discovered through periodic Bluetooth scans and location information from cell tower IDs.
Several other groups have performed similar studies. Some
 have used Bluetooth to measure device connectivity {\cite{realityMining}, \cite{haggle}, \cite{WR}}, while others rely on WiFi
{\cite{dartmouth}, GPS \cite{Rhee2007,Rhee2008,Rhee2009}, or the position of cell towers \cite{marta}.
The duration of experiments has varied from 2 days to over one year, and the numbers of
participants has also varied from $\sim 10$ to $\sim 100,000$.

It has been suggested that the probability density function (pdf) $p(t)$ of times between human contact is well approximated by a
truncated power law i.e. $p(t) \sim t^{-(1+\alpha)}$ over some range. This is so whether the contact is by physical proximity (i.e., detectability of wireless access points or Bluetooth devices, or closeness of GPS locations \cite{psn-HUMAN}, \cite{MSR-mobicom}, \cite{Rhee2007}) or by telecommunication (i.e., mobile phone call \cite{marta} or e-mail \cite{Malmgren2008}), and whether one or both contacting devices are in motion (e.g., both Bluetooth, one Bluetooth and fixed wireless access points, mobile phone and
fixed masts).

A summary is given in Table~I of studies in which the stability exponent $\alpha$ has been inferred
from an inter-contact time (ICT) distribution, together with the approximate range of
applicability. From the quoted values, $\alpha$ is inferred to be in the interval $[\approx 0,
0.9]$ which is within the allowable range ($0<\alpha \le 2$) for the tails of a L\'{e}vy (stable)
distribution \cite{Paul1999,Mantegna2000} (except possibly for the marginal case of the Europe
study of mobile phone contact which could actually be a gamma distribution.) Consequently it has
been argued that human mobility patterns resemble truncated L\'{e}vy walks (TLW). The TLW paradigm
represents a development of the L\'{e}vy flight, which was a random walk comprising steps drawn
from a L\'{e}vy distribution, rather than a Gaussian as occurs in the more familiar Brownian random
walks \cite{Shlesinger1995}.  The first modification, to a finite constant velocity, was dubbed a
L\'{e}vy walk. Subsequently the limitation to a finite domain was described as truncation
\cite{Mantegna2000}. More recently some researchers have also considered the velocity to be a
variable (e.g. \cite{Rhee2008}).

Similar movement patterns have also been inferred for animals \cite{Viswanathan1996}, and it has
been proposed that L\'{e}vy foraging is an optimal strategy under at least some circumstances
\cite{Viswanathan1999}. Debate continues as to the extent to which a L\'{e}vy strategy could be
universal and insensitive to the details of the environment and of the physiology and motivation of
the individual (e.g. \cite{Edwards2007,Reynolds2009}, and references therein). However, the
statistical analysis methods which have most frequently been used to infer empirical support for
the truncated L\'{e}vy walk hypothesis have recently been criticised, both in the ecology
literature and more generally \cite{Edwards2007,Edwards2008,Clauset2007,White2008}. Key problems
identified have included: (1) The widespread inference of  power law pdfs by the graphical method
of straight line fitting to histograms with double logarithmic axes; (2) the difficulty of
inferring power laws over very limited ranges; (3) the use of intrinsically  biased methods (such
as (1)) for estimating the power law exponent; and, perhaps most importantly,  (4) inadequate, or
even a complete lack of, alternative hypotheses.

In the light of this, it is worthwhile to consider how these problems might apply to the human
mobility studies cited above and summarised in Table~I. For example, some simply compare their
distributions with a straight line on a log-log plot with unavoidable bias and spread for the
inferred power law exponent \cite{psn-HUMAN,MSR-mobicom}. In addition, referring to Table~I, the
inference of a possible power law region is very weak for the Intel, Cambridge 1 and Toronto
experiments because the region is so limited ($\sim 1/3$ decade), presumably related to the small
samples ($\sim 1000$ contacts). The evidence is more convincing for the larger samples (INFOCOM
2005, UCSD, Dartmouth and Europe) with wider apparent power law regions. Alternative hypotheses to
the pure power law null model have been considered, such as the exponentially-truncated power law
\cite{MSR-mobicom,marta}, but only one study \cite{Rhee2009} has actually fitted and quantitatively
compared several alternative models to ICT distributions (albeit simulated), using the less biased
maximum likelihood estimate to infer the model parameters such as the power law exponent and Akaike
weights \cite{anderson1999} to compare the goodness of fits. Thus, at present, the inference of a
truncated power law ICT distribution directly from experiment is limited.

Indeed, it would be surprising if a truncated power law was a complete description of human ICT
considering our prior knowledge about the social habits and structures of humans, such as the
working day and family and community responsibilities \cite{Malmgren2008}. In fact it has been
recognised that the ICT distribution is not stationary and changes with the time of day
\cite{MSR-mobicom}. Spatial movement distributions also exhibit daily patterns \cite{marta} and
Fourier analyses of proximity edges have daily and weekly periodicities {\cite{realityMining}}. Similarly a
fundamental semi-diurnal periodicity was identified in an early study claiming a L\'{e}vy strategy
for animal foraging \cite{Viswanathan1996}. This suggests that alternative models combining
non-trivial randomness and periodic rhythms should be investigated. At present such more
complicated models are challenging to test rigorously (e.g., by MLE) but progress can nevertheless
be made by closer examination of the experimental ICT distribution using different graphical
methods and modelling.

In this paper we consider three similar human contact experiments of varying durations (section 2).
We analyse and model them to identify regularities that modify the underlying L\'{e}vy walk
behaviour (section 3). Then we briefly compare these analyses with others in the literature and
discuss how this hybrid behaviour may be modelled and will affect the efficiency of ad-hoc
communication (section 4).
\\
%
\section{Datasets}
\noindent
We analyse trace data from the Haggle project \cite{haggle} and Crawdad database {\cite{crawdad}},
collected using Bluetooth communication in a conference environment and two university study
environments. The configuration of data collection is summarised in
Table~II. ~~\\~~\\
{\bf MIT:} in the MIT Reality Mining project {\cite{realityMining}}, 100 smart phones were deployed
to students and staff at MIT over a period of 9 months. These phones were running software that
logged contacts.
\\
\noindent {\bf Cambridge 2:} in the Cambridge Haggle project {\cite{haggle}}, 36 iMotes (Intel Mote
ISN100-BA) were deployed to 1st year and 2nd year undergraduate students for 11 days to detect
proximity using Bluetooth. The iMote runs TinyOS and is equipped with an ARM7TDMI processor
operating at 12MHz, with 64kB of SRAM, 512kB of flash storage, and a multi-coloured LED, and a
Bluetooth 1.1 radio, which has a radio range around 30 meters.
\\
\noindent {\bf INFOCOM 2006:} also in the Cambridge Haggle project,
77 iMotes were deployed at the INFOCOM 2006 conference for 3 days.
\\

%
\begin{table}[t]
\begin{center}
\resizebox{!}{1.5cm}{
\begin{tabular}{|c|r|r|r|}
\hline
\scriptsize{Experimental data set} & \scriptsize{MIT} & \scriptsize{Cambridge 2} & \scriptsize{INFOCOM 2006}\\
\hline
\scriptsize{Device} & \scriptsize{Phone} & \scriptsize{iMote} & \scriptsize{iMote} \\
\scriptsize{Network type} & \scriptsize{Bluetooth} & \scriptsize{Bluetooth} & \scriptsize{Bluetooth}  \\
\scriptsize{Duration $L$ (days)} & \scriptsize{246} &   \scriptsize{11}  &   \scriptsize{3}    \\
\scriptsize{Granularity $\Delta$ (seconds)} & \scriptsize{600} &  \scriptsize{100}   &  \scriptsize{100}    \\
\scriptsize{Number of Devices} & \scriptsize{97}  &  \scriptsize{36} &  \scriptsize{77}  \\
\scriptsize{Number of Contacts} & \scriptsize{54,667}  &  \scriptsize{10,873} &  \scriptsize{191,336}  \\
\scriptsize{Average \# Contacts / Day} & \scriptsize{0.024}  &  \scriptsize{0.345} &  \scriptsize{6.7}  \\
\hline
\end{tabular}
}
{\bf \caption{{\bf Characteristics of the experiments}}}
\label{table:two}
\end{center}
\vspace{-6mm}
\end{table}
\begin{figure*}
\centering
\subfloat[][]{\includegraphics[height=6.8cm]{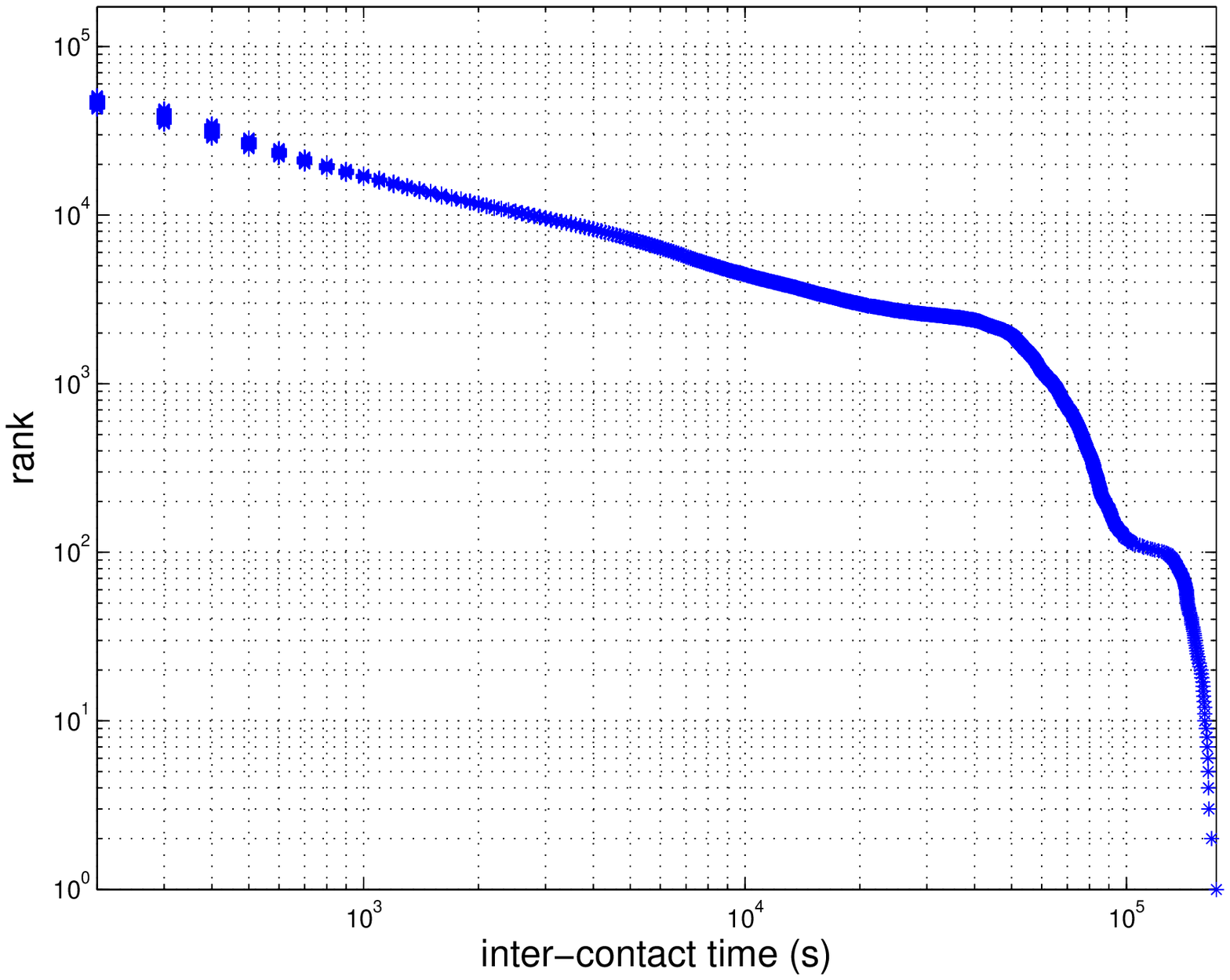}} \qquad
\subfloat[][]{\includegraphics[height=6.8cm]{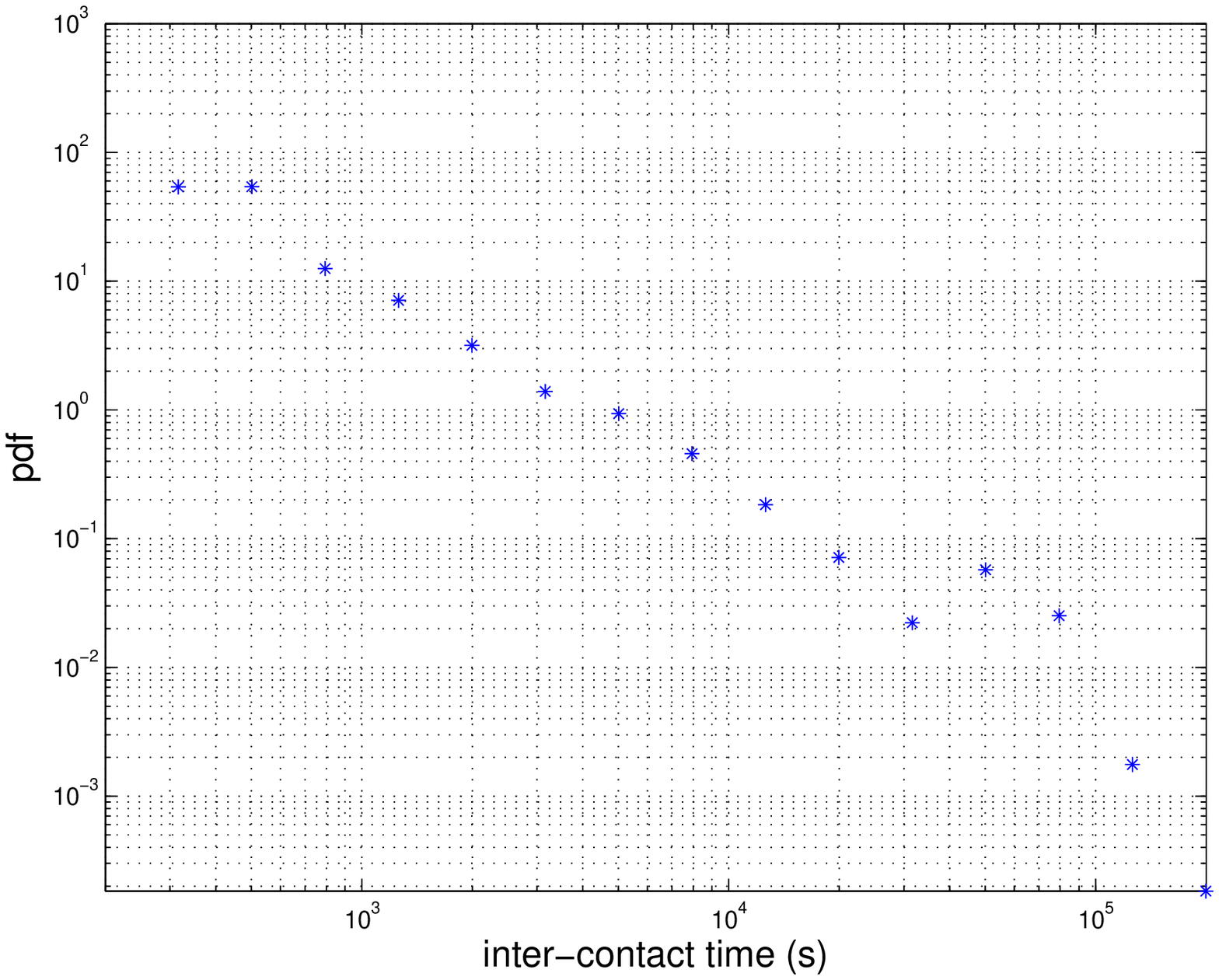}} \qquad
\subfloat[][]{\includegraphics[height=6.8cm]{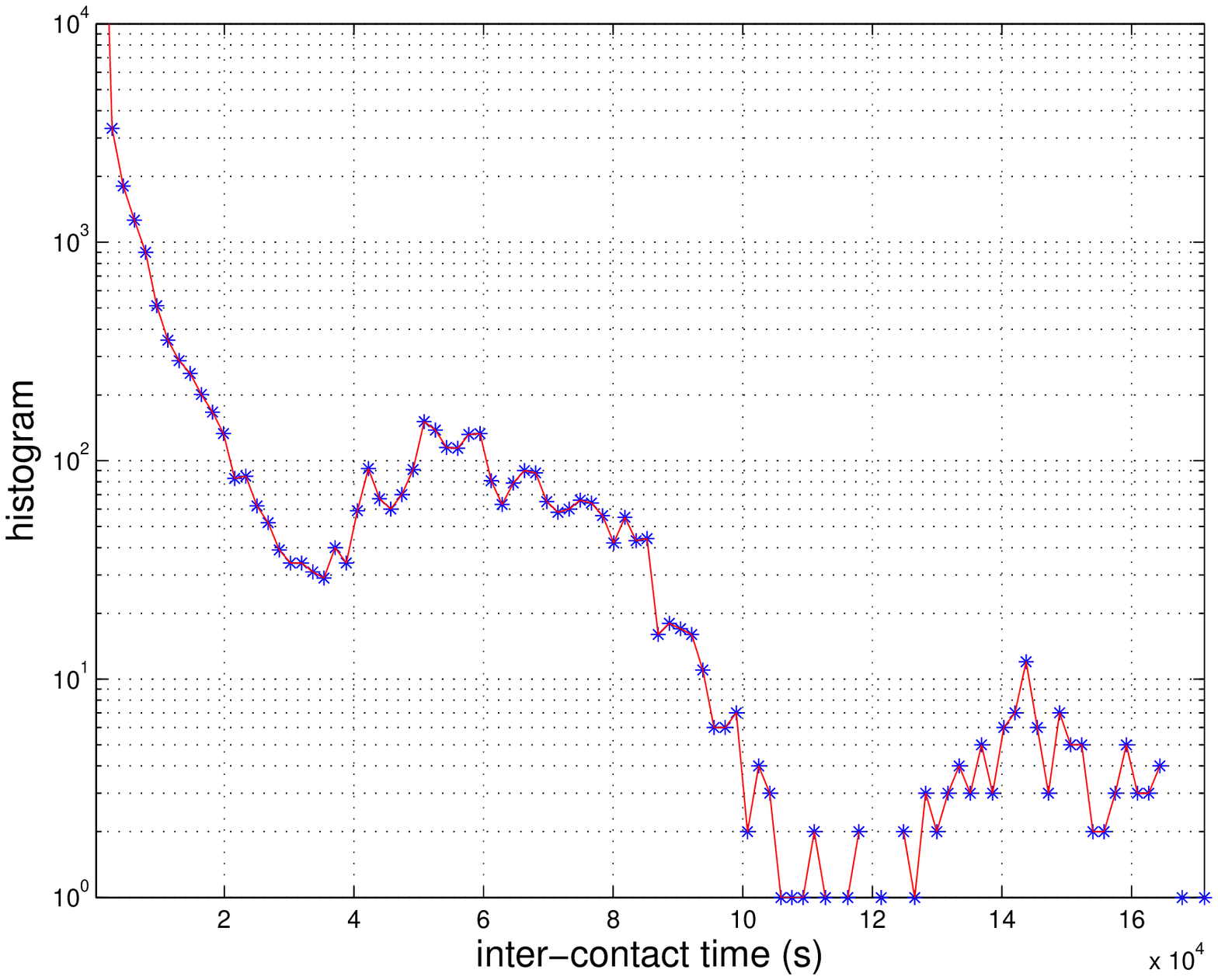}} \qquad
\subfloat[][]{\includegraphics[height=6.8cm]{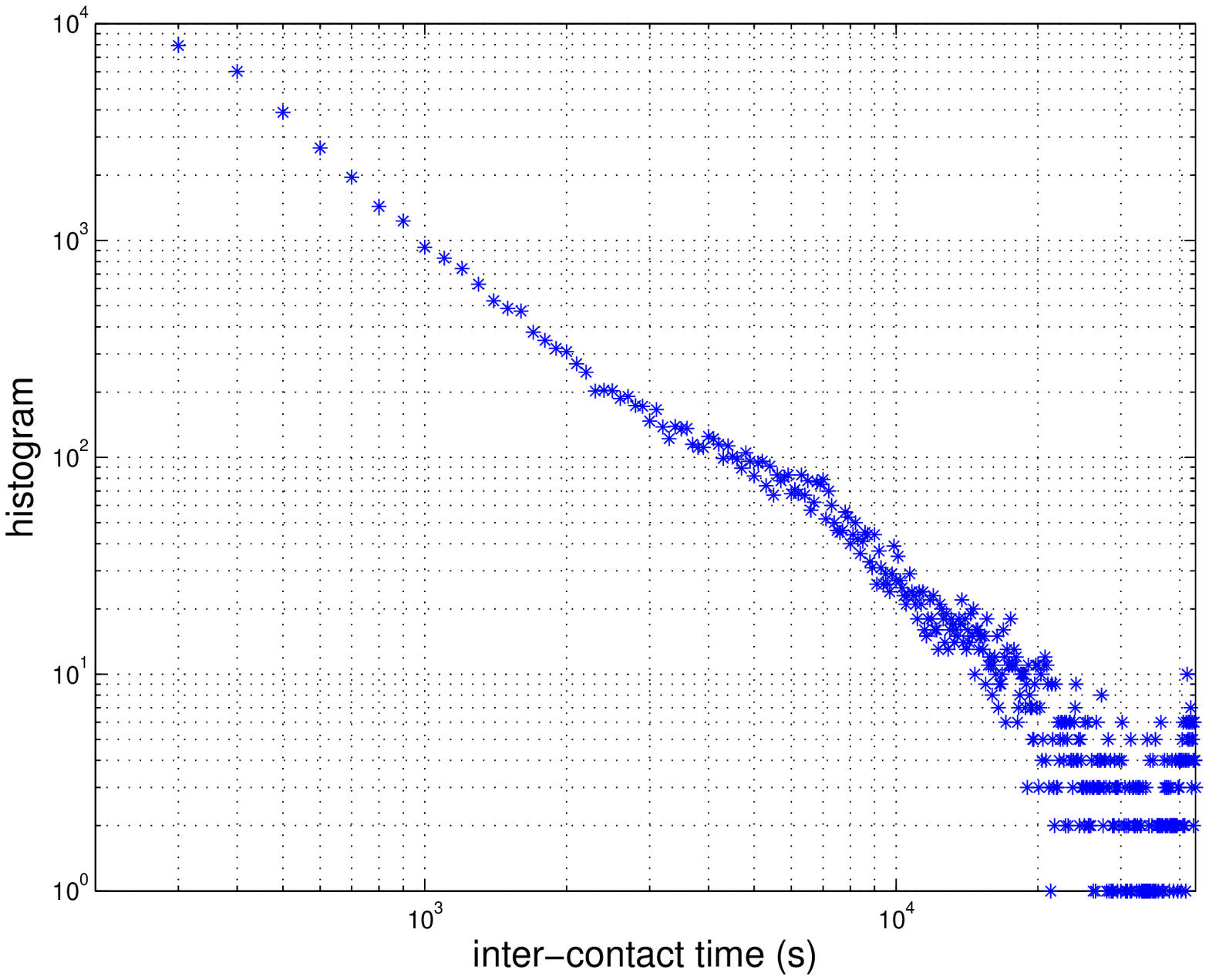}}
{\bf \caption{{\bf INFOCOM 2006: (a) Rank order plot,  (b) pdf, (c) semilog histogram with linear
bins, and (d) loglog histogram for times less than 12 hours, with linear bins}}} \vspace{-3mm}
\end{figure*}
\begin{figure*}
\centering \subfloat[][]{\includegraphics[height=6.8cm]{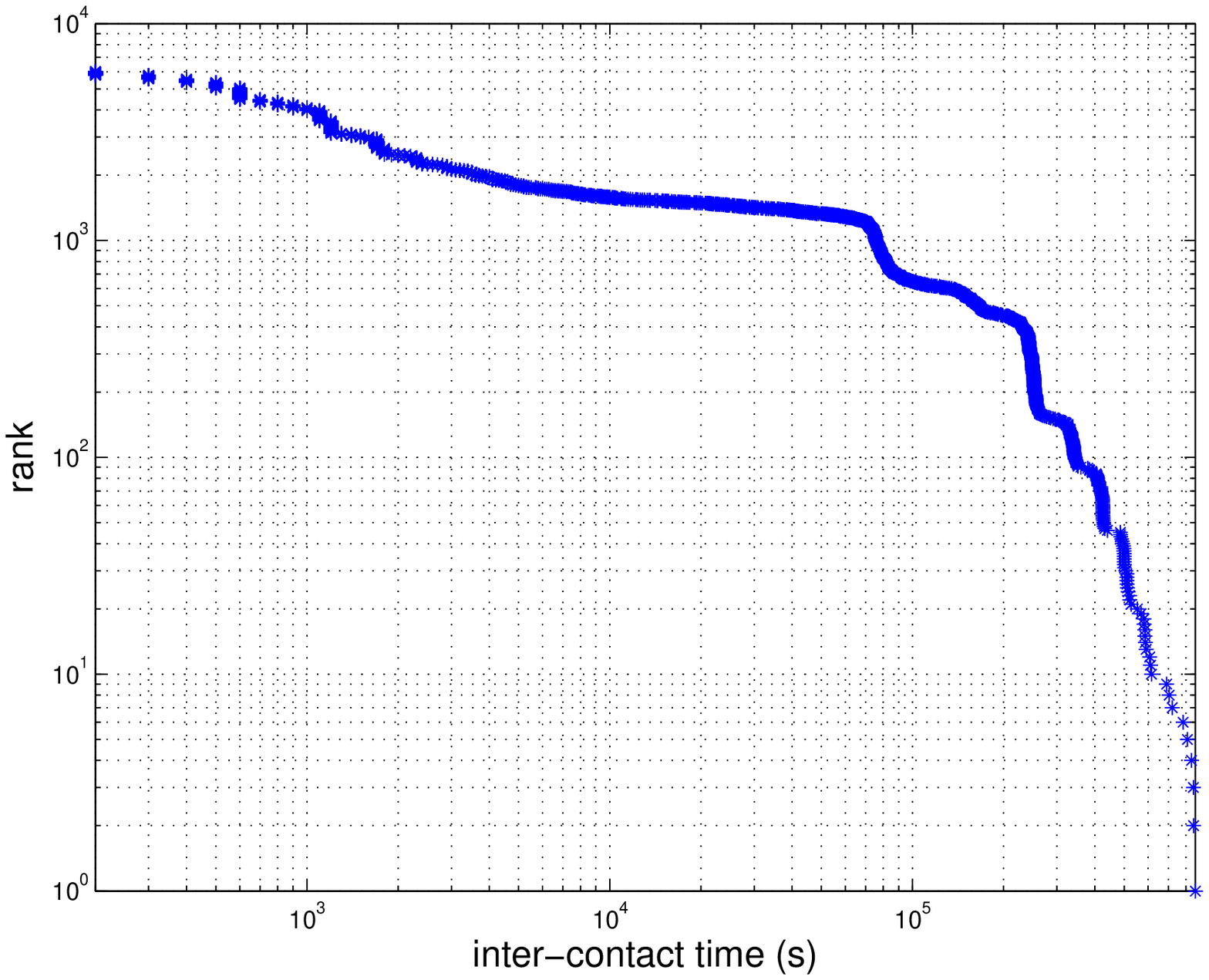}} \qquad
\subfloat[][]{\includegraphics[height=6.8cm]{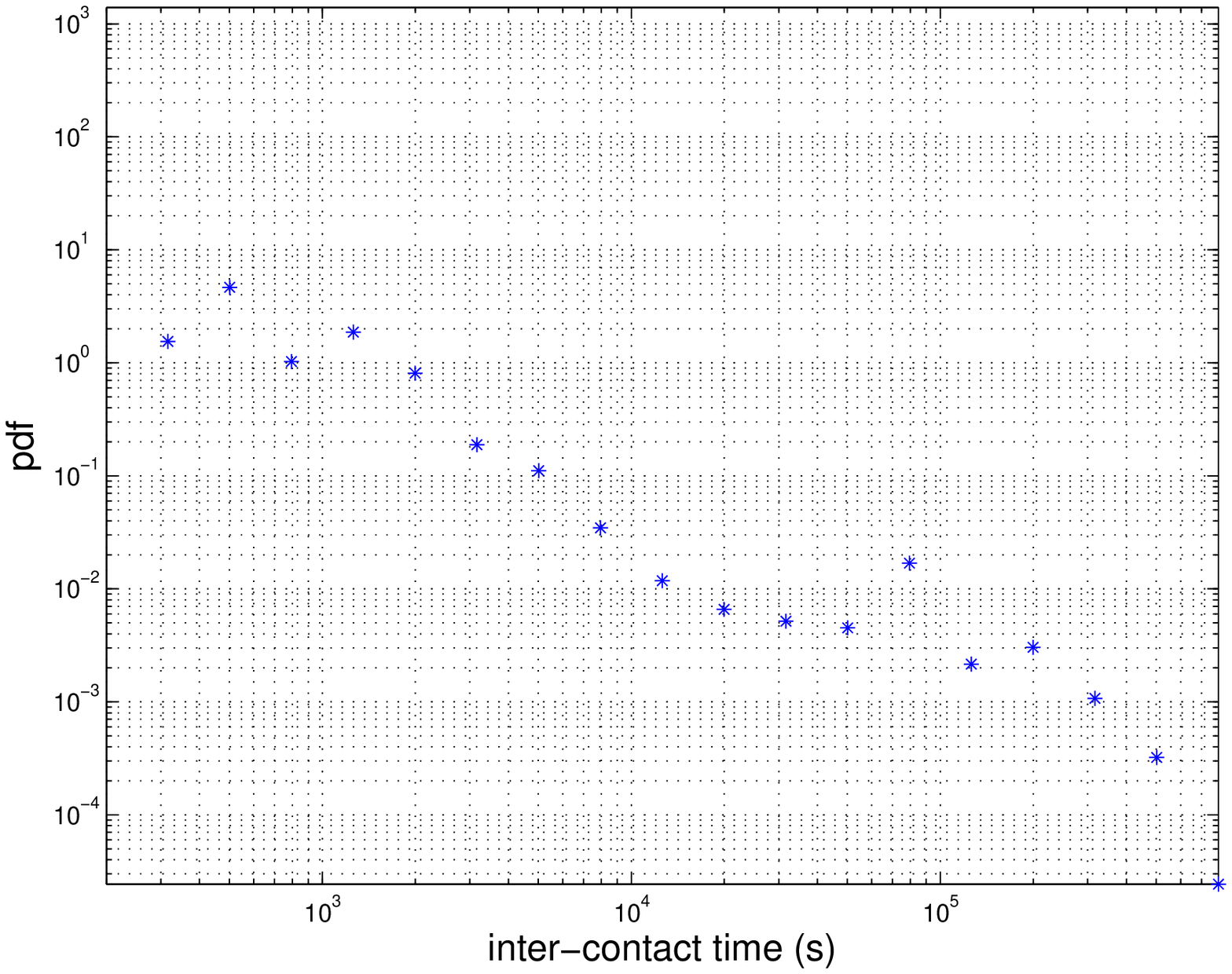}} \qquad
\subfloat[][]{\includegraphics[height=6.8cm]{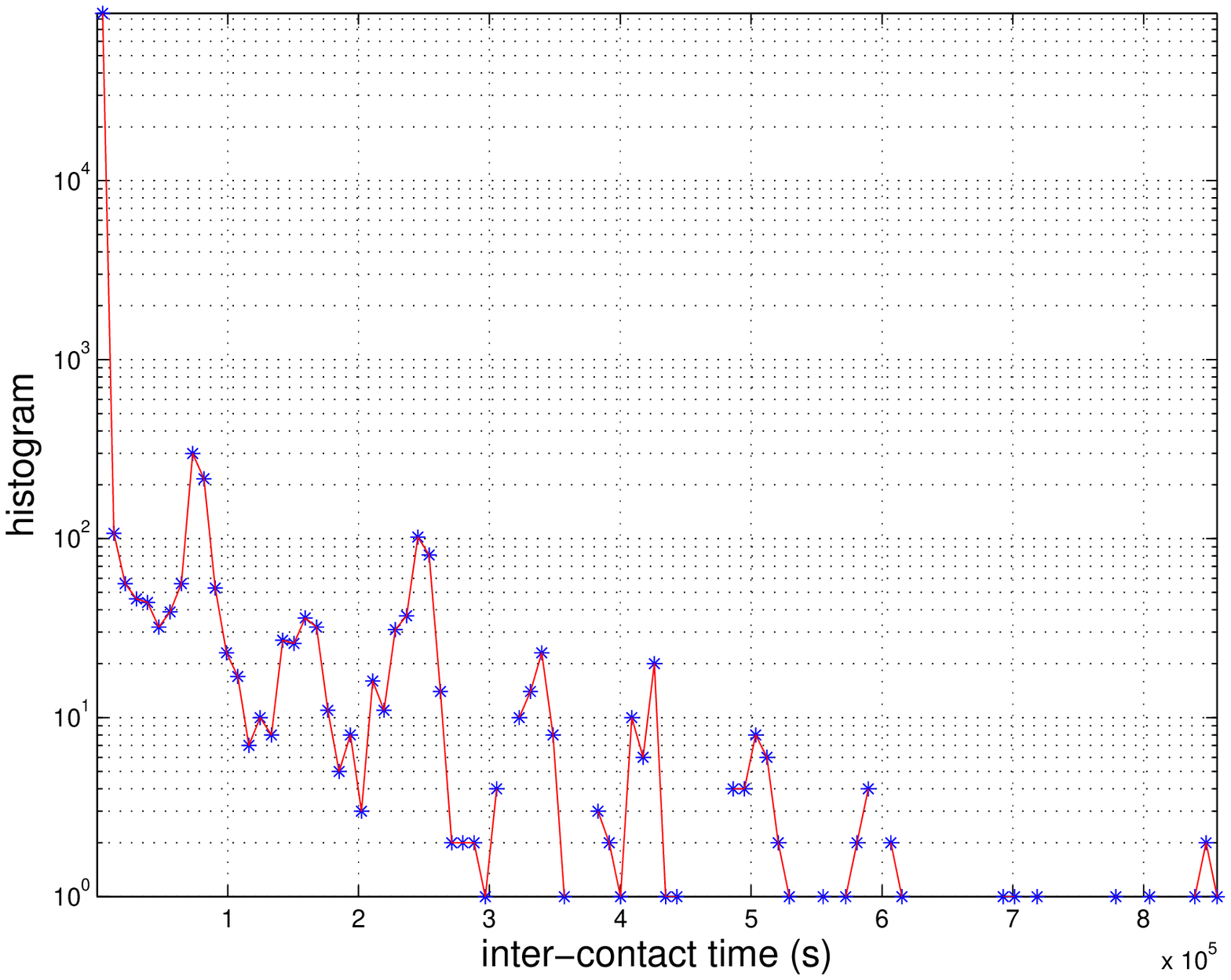}} \qquad
\subfloat[][]{\includegraphics[height=6.8cm]{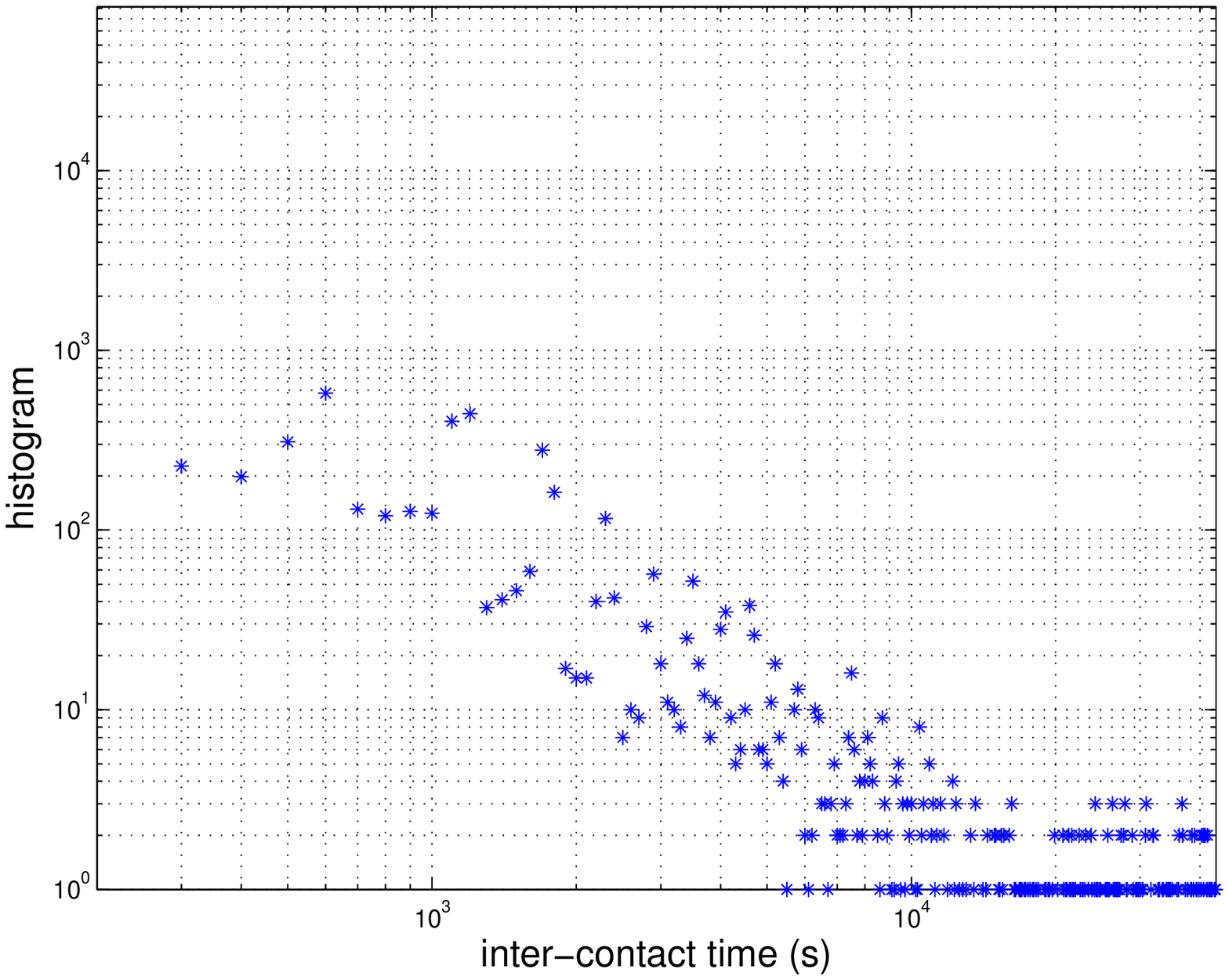}} {\bf \caption{{\bf Cambridge
2: (a) Rank order plot,  (b) pdf, (c) semilog histogram with linear bins, and (d) loglog histogram
for times less than 12 hours, with linear bins}}} \vspace{-3mm}
\end{figure*}

\begin{figure*}
\centering \subfloat[][]{\includegraphics[height=6.8cm]{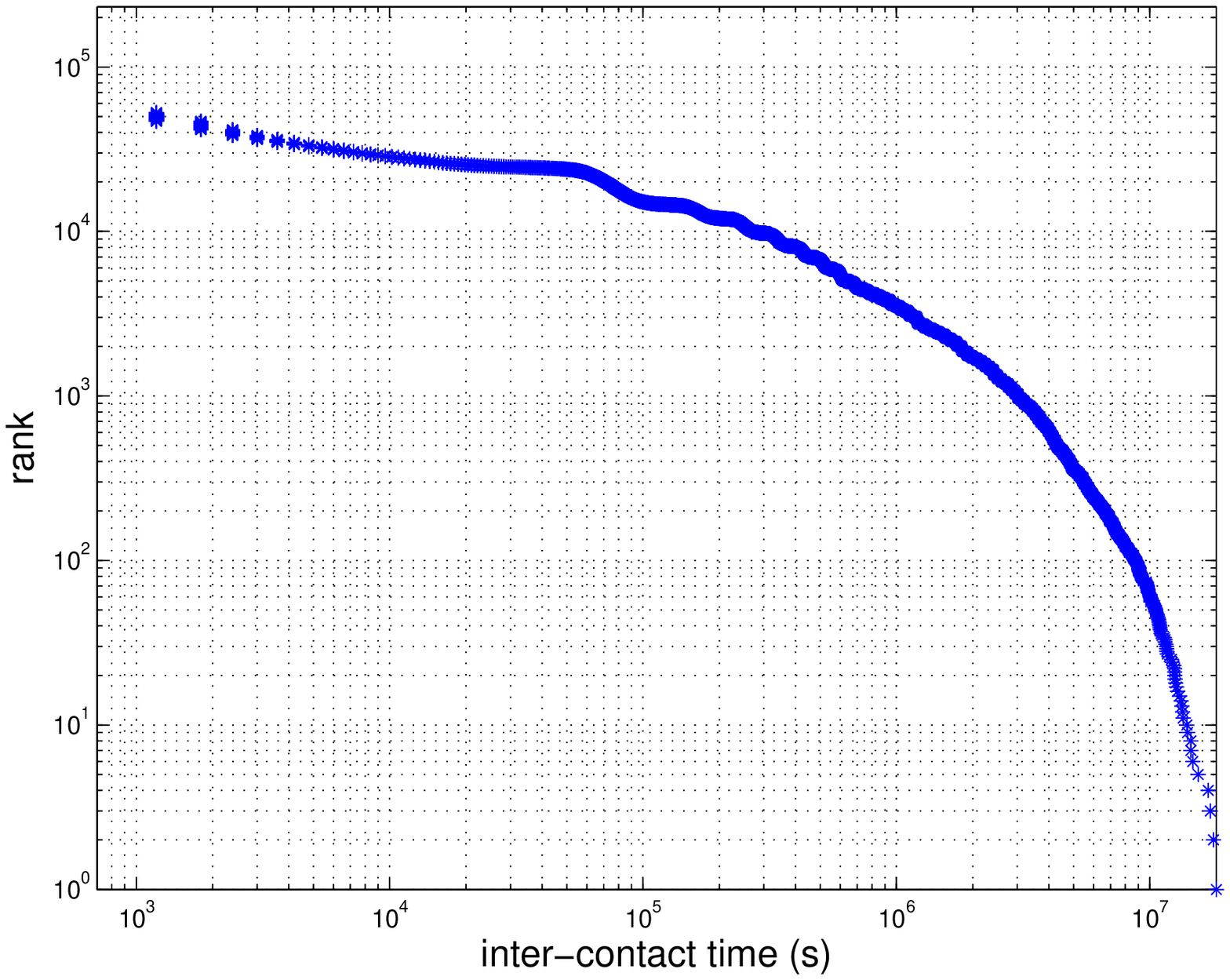}} \qquad
\subfloat[][]{\includegraphics[height=6.8cm]{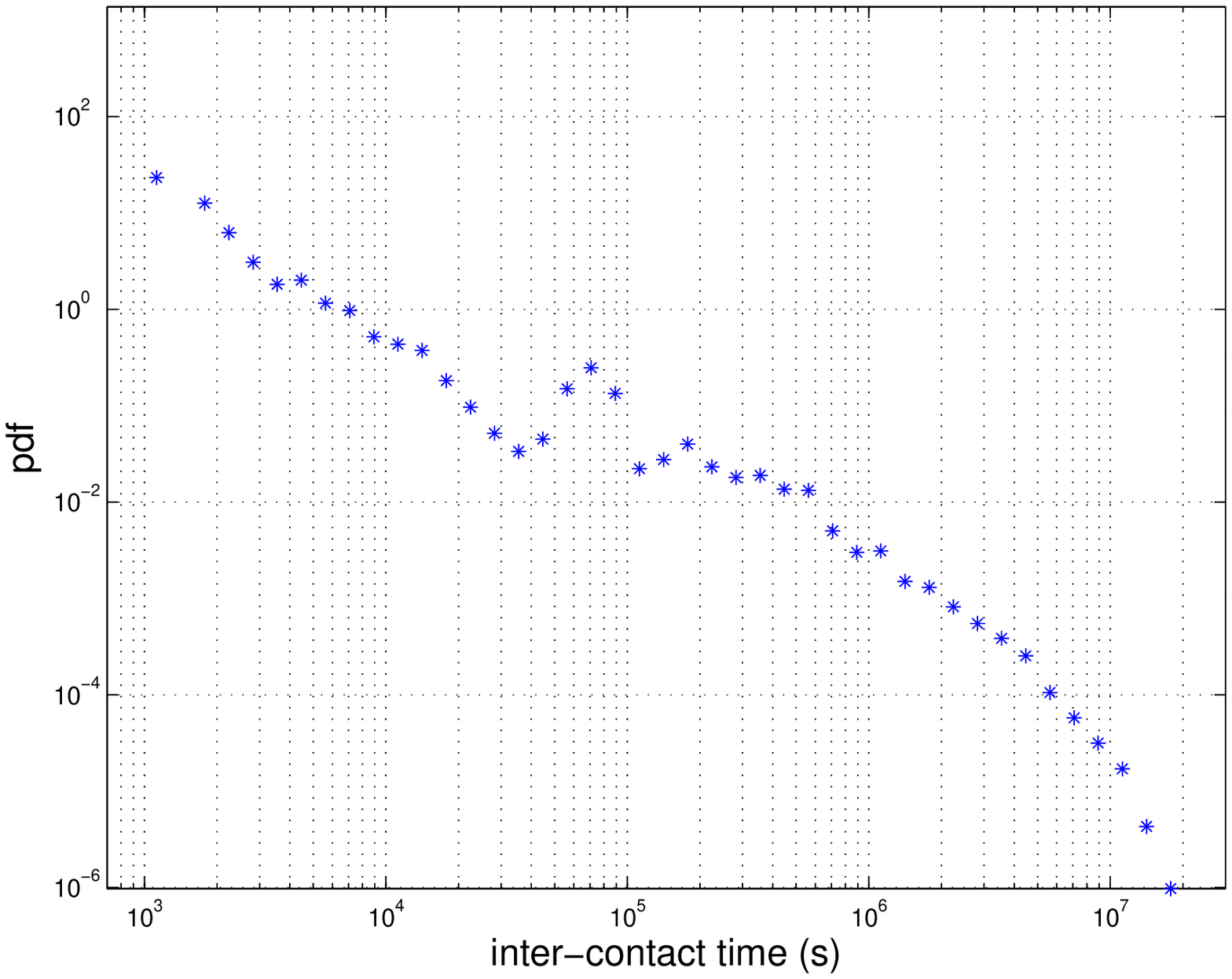}} \qquad
\subfloat[][]{\includegraphics[height=6.8cm]{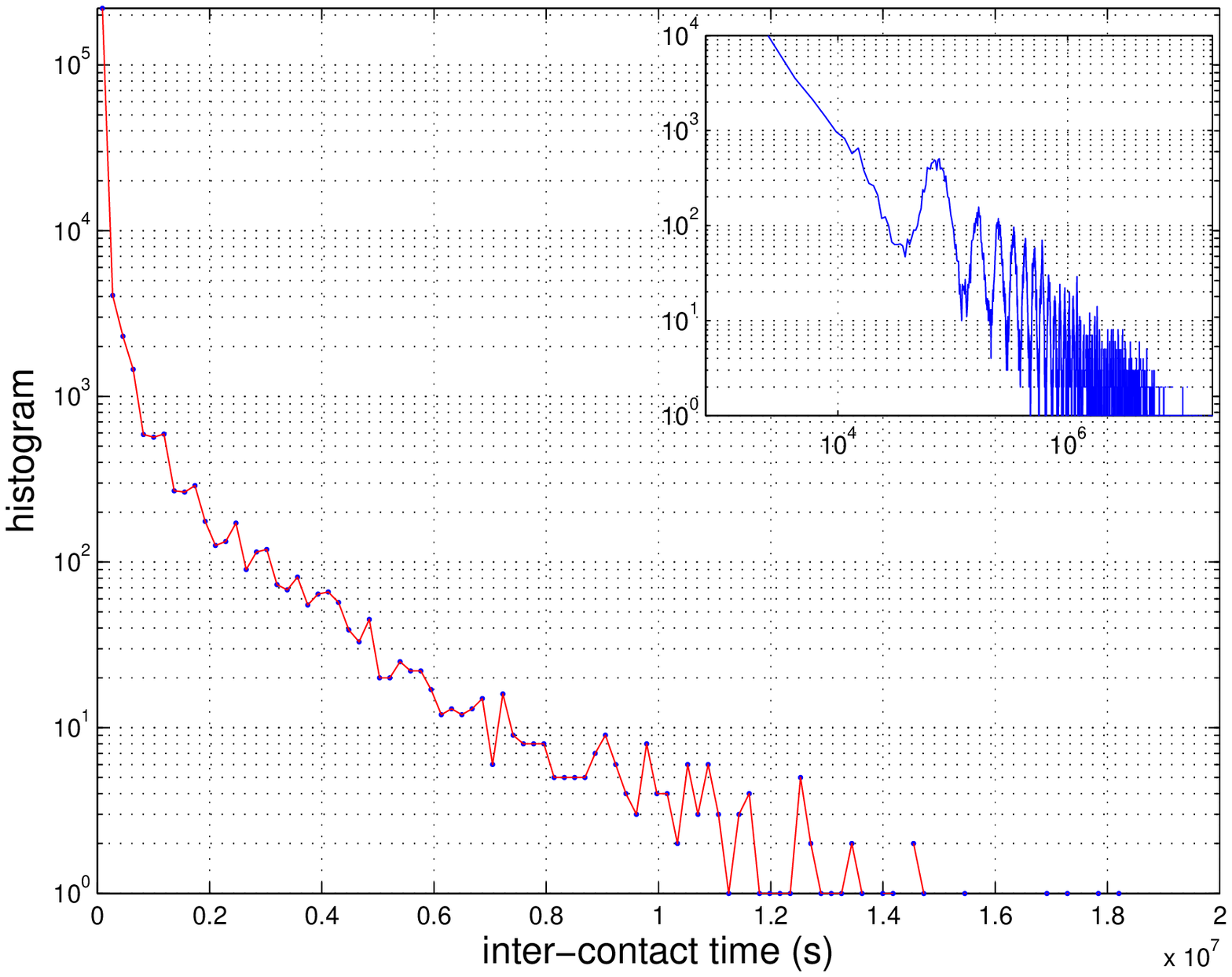}} \qquad
\subfloat[][]{\includegraphics[height=6.8cm]{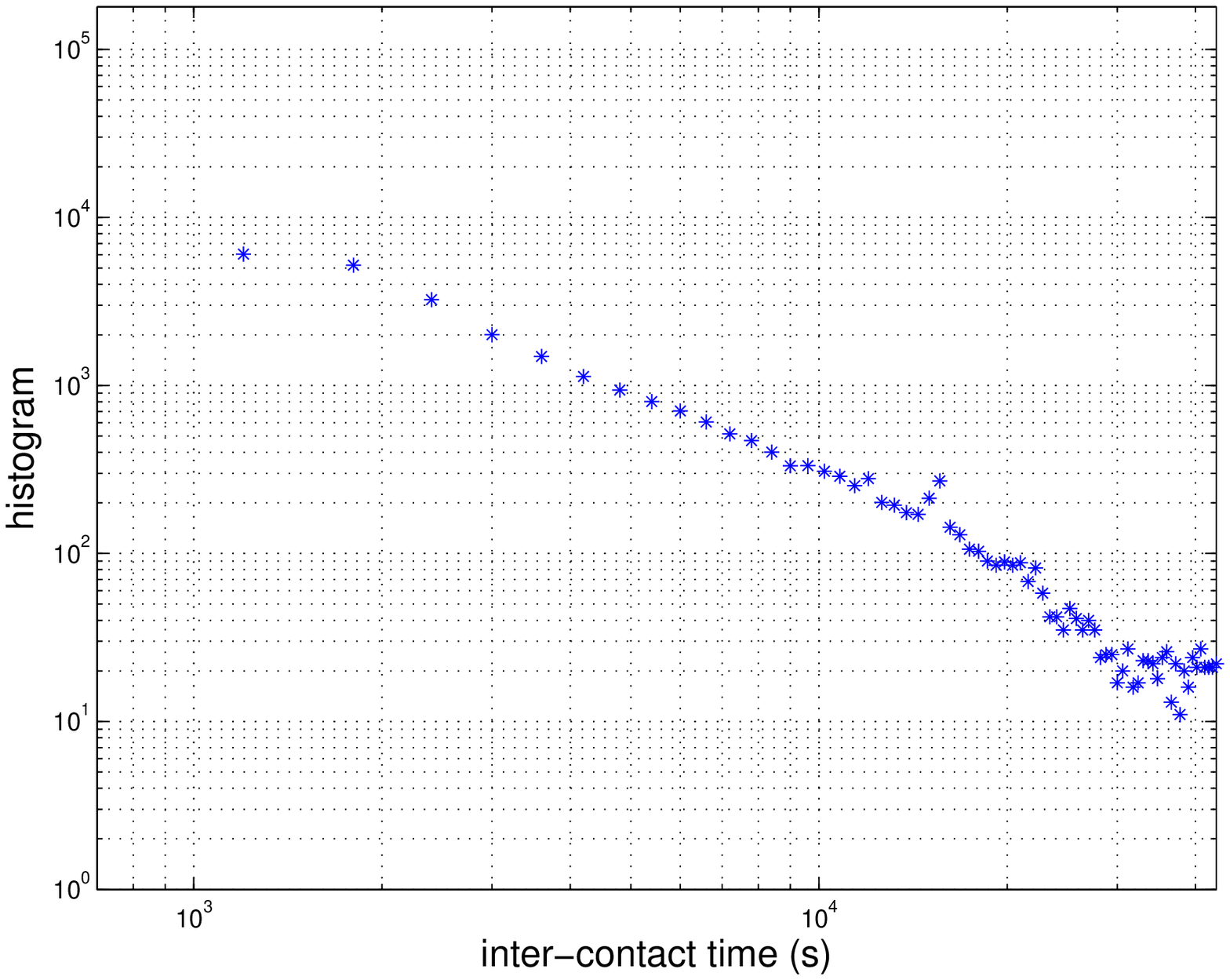}} {\bf \caption{{\bf MIT:
(a) Rank order plot, (b) pdf, (c) semilog histogram with linear bins (inset shows loglog histogram
with linear bins), and (d) loglog histogram for times less than 12 hours, with linear bins}}}
\vspace{-3mm}
\end{figure*}

The logged data from the above experimental studies are used to build
time-dependent network information to study the distribution of contact
times, inter-contact times, community structure and their statistical
properties, where we constructed discrete event traces of pair interactions
of 10 to 600 seconds intervals. We have aggregated raw data within 100 or
600 second time windows to avoid uncertainty of device detection from a
complex Bluetooth communication protocol.

A complex operation is required to collect accurate connectivity traces
using Bluetooth communication, as the device discovery protocol may limit
detection of the devices in radio proximity. Bluetooth uses a special
physical channel for devices to discover each other.
A device becomes discoverable by entering the inquiry substate where it can
respond to inquiries from other devices. The inquiry scan substate is used
to discover surrounding devices. The discovering device iterates (hops)
through all possible inquiry scan channel frequencies in a pseudo-random
fashion.
For each frequency, it broadcasts an inquiry and listens for responses.
Therefore, a Bluetooth
device cannot scan for other devices when the device cannot be in
discoverable. Bluetooth inquiry can only happen in $1.28$ second intervals.
It is reported that an interval of $4 \times 1.28 = 5.12$ seconds gives a
more than 90\% chance of finding a device. However, there is no available
data for situations where many devices are present, and no precise study has
been reported. The Bluetooth standard recommends being in the inquiry scan
substate for 10.24 seconds in order to collect all responses in an
error-free environment. A 10.24 seconds alternation may cause missing links,
and we therefore deploy 5.12 seconds for inquiry.
The power consumption of Bluetooth is also a critical limitation for the
scanning interval. The iMote connectivity traces in Haggle \cite{haggle} use
a scanning interval of approximately 2 minutes, while the Reality Mining
project in MIT \cite{realityMining}, with cell phones, uses 5 minutes. The
ratio of devices with Bluetooth enabled to the total number of devices is
around only an average 15\% - 20\% of population. The range of Bluetooth
varies between 10m and 80m, which depends on the device class such as cell
phones or laptops. In cell phones, the Bluetooth range is usually 5 - 10m.
We have observed that the devices can be detected in a 20m range if there
are no obstacles, while with obstacles such as a thick wall the range drops
to 5m (see more detail in \cite{WR}\cite{erc}).
\\


\section{Rhythm and randomness}
\noindent
In each of the experiments we calculated all possible inter-contact times $T$ between any two
nodes, where ICT is defined as the time between the end of contact between two nodes and the start
of next contact between the same two nodes. Figures 1-3 summarise the ICT distribution for the
three experiments. In each case, the distribution is plotted as (a) a rank order plot with double
logarithmic axes, (b) a probability density function with logarithmic co-ordinate (probability
density) axis and logarithmic ordinate (inter-contact time) axis using exponentially spaced bins
(i.e., equal bin width in logarithm space = 0.1 decade), (c) a histogram with logarithmic
co-ordinate (frequency) axis and linear ordinate (inter-contact time) axis using 100 equally-spaced
bins (equivalent to a pdf with linearly spaced bins to within a constant), and (d) a histogram for
inter-contact times up to 12 hours with logarithmic co-ordinate (frequency) axis and logarithmic
ordinate (inter-contact time) axis using equal bin widths at the granularity $\Delta = 100$~s or
600~s. (The inset in Figure~3c shows a double logarithmic histogram using equal bin widths of
1800~s.)
\\
\subsection{{\bf Truncated power law distribution}}
Considering the rank order plots in Figure 1-3, we might suggest as others have done that the ICT
tail distribution of all three experiments roughly resembles a restricted range power law with
exponent $< 1$ (cf Figure 1 and 2 of \cite{psn-HUMAN}). To illustrate this, we
performed the following simulation:

\begin{enumerate}
  \item A set of contact times is calculated for L\'{e}vy walks in a domain bounded by the duration of the experiment $L$
(see Table~II). Specifically we calculate the cumulative sum, $t_i = \sum_{j=1}^{i} X_j$, where $X$
is a set of $N$ iid samples chosen from the Pareto distribution with pdf $p(x) \sim x^{-(1+a)}$ in
the range $\Delta$ to $100 L$. The samples are generated by picking iid samples $C_i$ from the
uniform distribution in the range (0,1] and then inverting the analytical equation for the Pareto
cumulative probability distribution to find the value $x = X_i$ that yields the value $C = C_i$.
  \item Divide the contact times into individual trials (i.e., trial number $= t_i \hspace{.5em} \mbox{modulo} \hspace{.5em} L$).
  \item Calculate the set of inter-contact times $T$ from the time differences between neighbouring contact times, $T_i = t_{i+1} - t_i$, omitting inter-contact times that straddle trials.
\end{enumerate}

Figure~4a shows a simulated ICT probability
distribution (solid line) choosing $a = 0.4$ and other parameters corresponding to the
configuration of the INFOCOM 2006 experiment -- $\Delta = 100$~s, $L = 3$~days, and $N = 10,000$.
It is clear that the simulated distribution is only a crude approximation to the actual INFOCOM 2006 distribution
(dashed line) and other structure is evident.
\\
\subsection{{\bf Circadian rhythm}}
\begin{figure*}
\hspace{-10mm}
    \centering
    \begin{tabular}{c}
    \includegraphics[width=8.5cm]{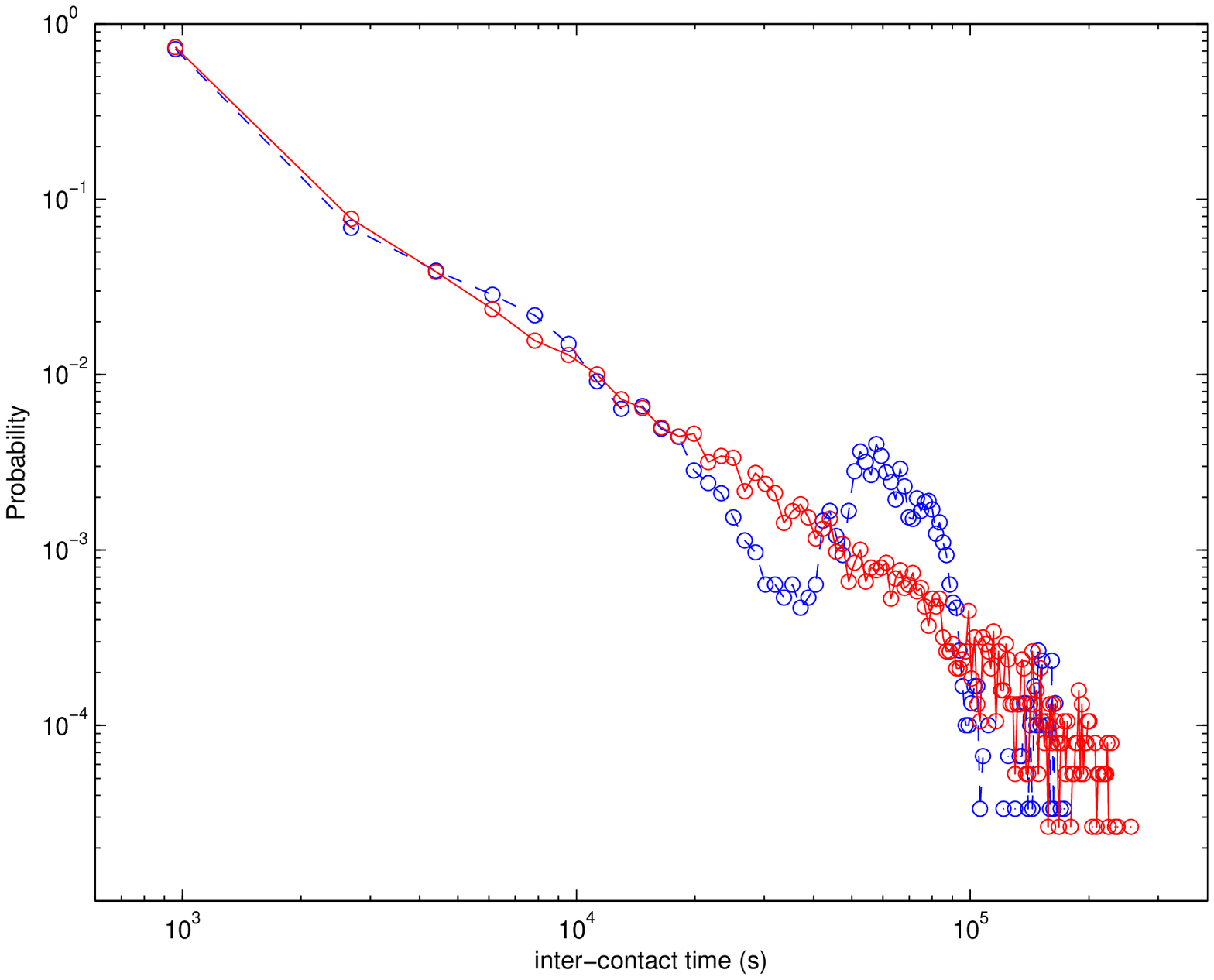}
    ~~~
    \includegraphics[width=8.5cm]{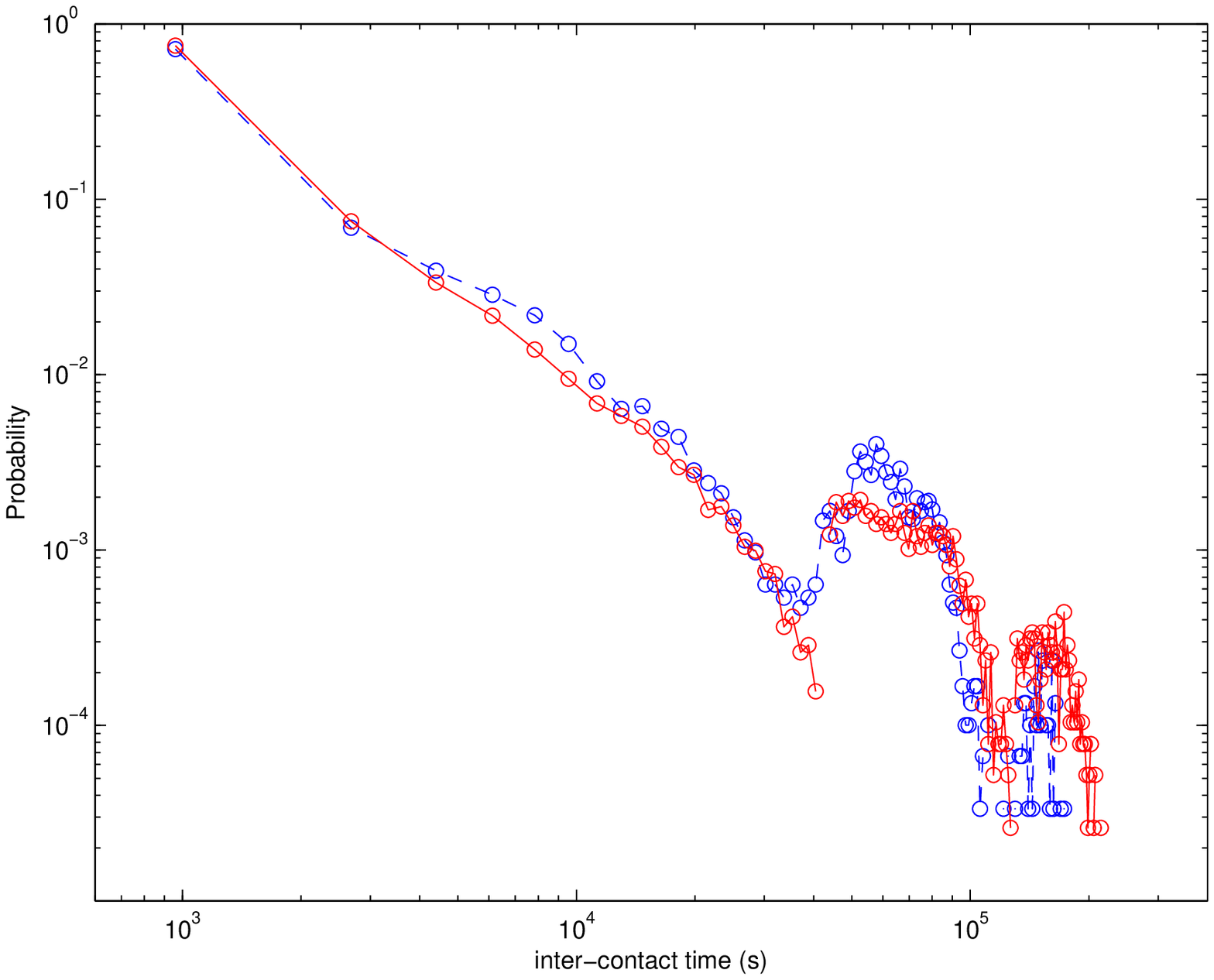}
    \end{tabular}
{\bf \caption{\label{periodical}{\bf Comparison of L\'{e}vy flight simulation of inter-contact times without (left), and with (right), the presence of circadian periodicity.  }}}
\vspace{-5mm}
\end{figure*}
This is also obvious in the other experiments (e.g., the histograms in figures 1c, 2c and 3c.)
where there are significant deviations about any candidate monotonic function. Closer inspection
reveals much of this deviation to be associated with a circadian rhythm, as evidenced by the
alignment of peaks in the histogram/PDF at integer multiples of 24 hours. (Note also a weekly rhythm
in Figure~3.)

Nevertheless, the INFOCOM 2006 and MIT distributions are well approximated by a power law on time scales much
less than a day (e.g., $< 12$ hours, see Figure~1d and 3d). (This is less obvious in the Cambridge 2 experiment
(Figure 2d) due to a $\approx 10$~min periodicity which is likely an experimental artefact). This suggests that a better
null model of ICT in these experiments is a
L\'{e}vy walk in a periodic domain. To investigate this we performed the following simulation:
\\
\begin{enumerate}
  \item A set of contact times is calculated for L\'{e}vy walks in a domain bounded by the duration of the experiment $L$ (see Table~II). Specifically we calculate the cumulative sum, $t_i = \sum_{j=1}^{i} X_j$, where $X$ is a set of $N$ iid samples chosen from the Pareto distribution with pdf $p(x) \sim x^{-(1+a)}$ in the range $\Delta$ to $L$. The samples are generated by picking iid samples $C_i$ from the uniform distribution in the range (0,1] and then inverting the analytical equation for the Pareto cumulative probability distribution to find the value $x = X_i$ that yields the value $C = C_i$.
  \item Divide the contact times into days and retain only contact times that fall within a working day, defined to start at $t_s$~h and end at $t_e$~h (i.e., $t_s \le d_i = t_i \hspace{.5em} \mbox{modulo} \hspace{.5em} 24 \mbox{hours} \le t_e$).
  \item Divide the contact times into individual trials (i.e., trial number $= t_i \hspace{.5em} \mbox{modulo} \hspace{.5em} L$).
  \item Calculate the set of inter-contact times $T$ from the time differences between neighbouring contact times, $T_i = t_{i+1} - t_i$, omitting inter-contact times that straddle trials.
\end{enumerate}
Figure~4b shows a simulated ICT probability distribution (solid line) choosing $a = 0.4$ and other parameters corresponding to the configuration of the INFOCOM 2006 experiment: $\Delta = 100$~s, $L = 3$~days, and $N = 10,000$. The
simulated distribution compares favourably with the actual INFOCOM 2006 distribution (dashed line),
supporting the null model over a L\'{e}vy walk confined within the domain $L$ but not within the working day.

\section{Conclusions and Implications}
\noindent
The distribution of human inter-contact times from three experiments of differing durations has
been analysed using different graphical presentations. This has revealed three essential properties
of human contact:
\\\\
{\it \bf Random, scale-free.} On sufficiently short time scales, the ICT distribution is
approximated by a power law consistent with the return times of a L\'{e}vy flight. The value of the
stability exponent ($\alpha < 1$) implies no characteristic ICT in the absence of other
constraints.
\\\\
{\it \bf Truncated.} At some time scale the power law component is truncated by a constraint on
inter-contact time. One artificial constraint is the experiment itself which prohibits recording
ICTs longer than the experiment duration. This is demonstrated in the simulated ICT distribution in
Figure 4a and should be considered in comparing results from experiments of differing durations.
More significantly, another constraint is the removal of agents from the contact domain. An example
of this is movement from work to home which suppresses ICTs between agents in the same work group
on times scales beyond the working day. This is demonstrated in the simulated ICT distribution in
Figure 4b by the truncation of the power law component at $ICT \sim 10^4$~s.
\\\\
{\it \bf Periodic.} Environmental, biological, and social constraints may have rhythms that
encourage repeated encounters such as the daily to-ing and fro-ing between work and home. This is
demonstrated in the simulated ICT distribution in Figure 4b by the peak at $ICT \sim 6 \times
10^4$~s and $\sim 15 \times 10^4$~s (i.e., separated by 24 hours).
\\\\
These three properties have been previously surmised by various different means but evidence of
their co-existence in the ICT distribution has been overlooked. In particular, closer examination
of previously published ICT distributions (e.g., \cite{marta}) reveals deviations about a truncated
power law consistent with a circadian rhythm. Recognition of this rhythm in the empirical
distribution is important otherwise models of human movement and behaviour may be unrealistically
modified to generate only the scale-free property (e.g., \cite{Boldrini2009}). It also has
significant implications for building efficient routing algorithms and functionality on top of
opportunistic networks. As a very simple example, clearly a rhythm of period $P$ that removes
agents from each other for a time $P/2$ reduces the average number of contacts by 50\% over
multiple cycles. But its determinism might also be exploited to increase communication efficiency.
For example, the time of the next encounter could be estimated at the node and thus selection of
the next hop could be determined based on the expected shortest time to the next encounter. The
periodic behaviour of nodes could indicate moving from one network partition to another and this
could be used for temporal clustering of nodes, where temporal-based communities could be used as a
backbone of logical network structure for forwarding \cite{mobihoc}. By these means,
mobility-assisted forwarding can take advantage of patterns arising in the distribution of nodes in
time and space. One alternative movement model is suggested that combines the L\'{e}vy walk model
with models such as the Home Cell Mobility Model (e.g., \cite{Boldrini2009}) that incorporate the
influence of social structure. However the development of more complicated models will also present
challenges in testing them and distinguishing between competing models.
%
%
\section*{Acknowledgement}
\noindent
The research is part funded by the EU grants for the Haggle project, IST-4-027918, the SOCIALNETS
project, 217141, and the EPSRC DDEPI Project, EP/H003959. MPF and NWW are supported by the Natural
Environment Research Council.
\bibliographystyle{ieee}

\end{document}